\documentclass[aps,prl,reprint,floatfix,longbibliography,superscriptaddress,tightenlines,twocolumn]{revtex4-2}
\usepackage{graphicx}
\usepackage{times}
\usepackage{dcolumn}
\usepackage{bm}
\usepackage{amsmath}
\usepackage[colorlinks,linkcolor=blue,anchorcolor=blue,citecolor=blue]{hyperref}
\newcommand{\rfig}[1]{Fig.~\ref{#1}}

\newcommand{\req}[1]{Eq.~(\ref{#1})}
\usepackage[T1]{fontenc}

\begin{document}

\title{High-order topological pumping on a superconducting quantum processor}
\author{Cheng-Lin Deng}
\thanks{These authors contributed equally to this work.}

\affiliation{Institute of Physics, Chinese Academy of Sciences, Beijing 100190, China}
\affiliation{School of Physical Sciences, University of Chinese Academy of Sciences, Beijing 100049, China}

\author{Yu Liu}
\thanks{These authors contributed equally to this work.}

\affiliation{Institute of Physics, Chinese Academy of Sciences, Beijing 100190, China}
\affiliation{School of Physical Sciences, University of Chinese Academy of Sciences, Beijing 100049, China}
\thanks{These authors contributed equally to this work.}
\author{Yu-Ran Zhang}
\thanks{These authors contributed equally to this work.}
\affiliation{School of Physics and Optoelectronics, South China University of Technology, Guangzhou 510640, China}

\author{Xue-Gang Li}
\affiliation{Beijing Academy of Quantum Information Sciences, Beijing 100193, China}

\author{Tao Liu}
\affiliation{School of Physics and Optoelectronics, South China University of Technology, Guangzhou 510640, China}
\author{Chi-Tong Chen}
\affiliation{Institute of Physics, Chinese Academy of Sciences, Beijing 100190, China}
\affiliation{School of Physical Sciences, University of Chinese Academy of Sciences, Beijing 100049, China}

\author{Tong Liu}
\affiliation{Institute of Physics, Chinese Academy of Sciences, Beijing 100190, China}
\affiliation{School of Physical Sciences, University of Chinese Academy of Sciences, Beijing 100049, China}

\author{Cong-Wei Lu}
\affiliation{Department of Physics, Applied Optics Beijing Normal University, Beijing 100875, China}

\author{Yong-Yi Wang}
\affiliation{Institute of Physics, Chinese Academy of Sciences, Beijing 100190, China}
\affiliation{School of Physical Sciences, University of Chinese Academy of Sciences, Beijing 100049, China}

\author{Tian-Ming Li}
\affiliation{Institute of Physics, Chinese Academy of Sciences, Beijing 100190, China}
\affiliation{School of Physical Sciences, University of Chinese Academy of Sciences, Beijing 100049, China}

\author{Cai-Ping Fang}
\affiliation{School of Physical Sciences, University of Chinese Academy of Sciences, Beijing 100049, China}

\author{Si-Yun Zhou}
\affiliation{School of Physical Sciences, University of Chinese Academy of Sciences, Beijing 100049, China}

\author{Jia-Cheng Song}
\affiliation{School of Physical Sciences, University of Chinese Academy of Sciences, Beijing 100049, China}

\author{Yue-Shan Xu}
\affiliation{Institute of Physics, Chinese Academy of Sciences, Beijing 100190, China}
\affiliation{School of Physical Sciences, University of Chinese Academy of Sciences, Beijing 100049, China}

\author{Yang He}
\affiliation{Institute of Physics, Chinese Academy of Sciences, Beijing 100190, China}
\affiliation{School of Physical Sciences, University of Chinese Academy of Sciences, Beijing 100049, China}

\author{Zheng-He Liu}
\affiliation{Institute of Physics, Chinese Academy of Sciences, Beijing 100190, China}
\affiliation{School of Physical Sciences, University of Chinese Academy of Sciences, Beijing 100049, China}

\author{Kai-Xuan Huang}
\affiliation{Beijing Academy of Quantum Information Sciences, Beijing 100193, China}

\author{Zhong-Cheng Xiang}
\affiliation{Institute of Physics, Chinese Academy of Sciences, Beijing 100190, China}
\affiliation{School of Physical Sciences, University of Chinese Academy of Sciences, Beijing 100049, China}
\affiliation{Hefei National Laboratory, Hefei 230088, China}

\author{Jie-Ci Wang}
\affiliation{Department of Physics and Key Laboratory of Low Dimensional\\ Quantum Structures and Quantum Control of Ministry of Education, Hunan Normal University, Changsha, Hunan 410081, China}
 
\author{Dong-Ning Zheng}
\affiliation{Institute of Physics, Chinese Academy of Sciences, Beijing 100190, China}
\affiliation{School of Physical Sciences, University of Chinese Academy of Sciences, Beijing 100049, China}
\affiliation{Songshan Lake Materials Laboratory, Dongguan, Guangdong 523808, China}
\affiliation{CAS Center for Excellence in Topological Quantum Computation, UCAS, Beijing 100190, China, and Mozi Labratory, Zhengzhou 450007, China}
\affiliation{Hefei National Laboratory, Hefei 230088, China}

\author{Guang-Ming Xue}
\affiliation{Beijing Academy of Quantum Information Sciences, Beijing 100193, China}
\affiliation{Hefei National Laboratory, Hefei 230088, China}

\author{Kai Xu}
\email{kaixu@iphy.ac.cn}
\affiliation{Institute of Physics, Chinese Academy of Sciences, Beijing 100190, China}
\affiliation{Beijing Academy of Quantum Information Sciences, Beijing 100193, China}
\affiliation{Songshan Lake Materials Laboratory, Dongguan, Guangdong 523808, China}
\affiliation{CAS Center for Excellence in Topological Quantum Computation, UCAS, Beijing 100190, China, and Mozi Labratory, Zhengzhou 450007, China}
\affiliation{Hefei National Laboratory, Hefei 230088, China}
\author{H. F. Yu}
\email{hfyu@baqis.ac.cn}
\affiliation{Beijing Academy of Quantum Information Sciences, Beijing 100193, China}

\author{Heng Fan}
\email{hfan@iphy.ac.cn}
\affiliation{Institute of Physics, Chinese Academy of Sciences, Beijing 100190, China}
\affiliation{Beijing Academy of Quantum Information Sciences, Beijing 100193, China}
\affiliation{Songshan Lake Materials Laboratory, Dongguan, Guangdong 523808, China}
\affiliation{CAS Center for Excellence in Topological Quantum Computation, UCAS, Beijing 100190, China, and Mozi Labratory, Zhengzhou 450007, China}
\affiliation{Hefei National Laboratory, Hefei 230088, China}

\date{\today}

\begin{abstract}
High-order topological phases of matter refer to the systems of $n$-dimensional bulk with the topology of $m$-th order, exhibiting $(n-m)$-dimensional boundary modes
and can be characterized by topological pumping.
Here, we experimentally demonstrate two types of second-order topological pumps, forming four 0-dimensional corner localized states on a 4$\times$4 square lattice array of 16 superconducting qubits. The initial ground state of the system for half-filling, as a product of four identical entangled 4-qubit states, is prepared using an adiabatic scheme.
During the pumping procedure, we adiabatically modulate the superlattice Bose-Hubbard Hamiltonian by
precisely controlling both the hopping strengths and on-site potentials. 
At the half pumping period, the system evolves to a corner-localized state in a quadrupole configuration. The robustness of the second-order topological pump is also investigated by introducing different on-site disorder. Our work studies the topological properties of high-order topological phases from the dynamical transport picture using superconducting qubits, which would inspire further research on high-order topological phases.
\end{abstract}

\maketitle

\begin{figure*}
    \centering
    \includegraphics[width=0.77\textwidth]{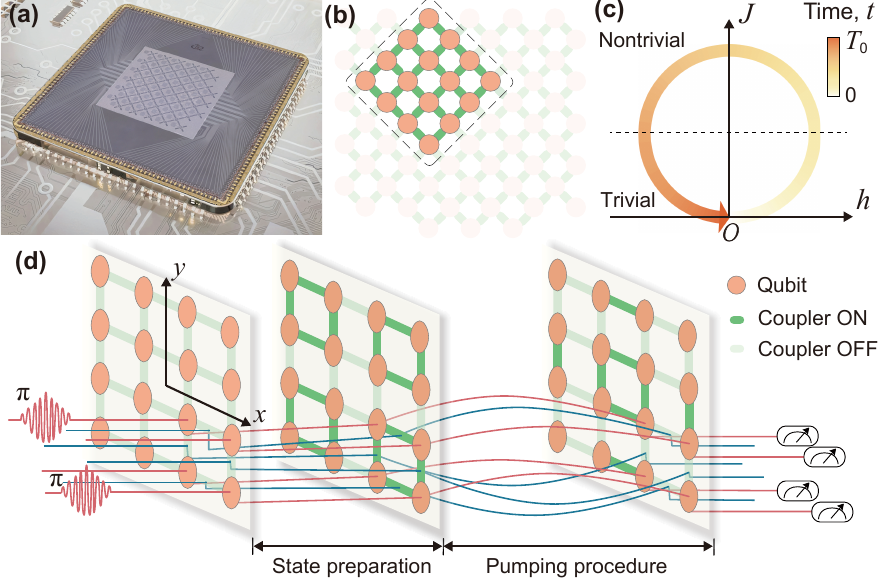}
    \caption{\label{fig:combfig1}
        {Quantum processor and experimental scheme.}
        (a) Photograph of the superconducting processor that is fabricated using the 
        flip-chip technique. (b) Qubits array. The processor is integrated with 62 qubits and 105 couplers, forming a square lattice. A subset of 4$\times$4 qubits and 24 couplers are used in our experiments.
        (c) Pumping trajectory. Parameters of the hopping strength $J$ and the on-site potential $h$ are dynamically modulated following a closed loop for a full pumping period.
        (d) Lattice configuration and experimental sequences. For each 4-qubit sub-lattice (down-right 4$\times$4 block as an example), the initial ground state of the system is prepared by firstly exciting the diagonal two qubits to states $|1\rangle $ and then applying adiabatic controls of the frequencies of both qubits and couplers. 
        During the pumping procedure, the system Hamiltonians evolve slowly by 
        tuning the hopping strengths and the on-site potentials between and on qubits, respectively.
        }
\end{figure*}

\emph{Introduction.}---High-order symmetry-protected topological (HOSPT) phases of matter derive from the electric dipole insulators to the multipole insulators in a modern formulation of Berry phase in band structures~\cite{benalcazar2017quantized,resta1998quantum}. 
The bulk-boundary correspondence for the quadrupole insulators establishes a connection between a two-dimensional bulk in a second-order topological phase and zero-dimensional topological-protected corner-localized states~\cite{citro2023thouless}.  
Recent literature reported the experiments on the HOSPT phases by
observing the topologically non-trivial corner states on platforms such as photons~\cite{benalcazar2022higherorder,mittal2019photonic,zilberberg2018photonic,wang2009observation,hafezi2013imaging,rechtsman2013photonic}, phonons~\cite{nash2015topological,susstrunk2015observation,serra2018observation}, electric circuits~\cite{ningyuan2015time,bao2019topoelectrical}, and metamaterial of microwave resonators~\cite{peterson2018quantized,ni2020demonstration,imhof2018topolectrical}.
However, to demonstrate quantized charge transport of the HOSPT phases remains very elusive.
Recently,  Ref.~\cite{wienand2022thouless} shows that a topological pump, e.g., the Thouless pump~\cite{thouless1983quantization,niu1984quantised,nakajima_topological_2016, lohse2016thouless, lohse2018exploring, zilberberg_photonic_2018, cerjan2020thoulessa, nakajima2021competition, kao_science_2021,jurgensen_quantized_2021,benalcazar_higher-order_2022,tao_interaction-induced_2023, walter_quantization_2023, xiang_simulating_2023,liu_disorder-induced_2024}, can also provide a dynamical characterization of HOSPT phases, which depends on the topology of the pump cycle. Corner states appear during topological pumping on a 2D superlattice  Bose-Hubbard model with staggered hopping strengths. 
A tuple of four Chern numbers is defined to measure quantized charge transport for the $C_4$-symmetric HOSPT phases~\cite{wienand2022thouless}. 

Here, we experimentally demonstrate diagonal and non-diagonal HOSPT pumps
 on a 4$\times $4 square lattice array of superconducting qubits. The initial ground state of the Hamiltonian for half-filling, separable as a product of four 4-qubit entangled states,  is prepared using an adiabatic scheme.
 During the pumping process, we tune the frequencies of the couplers, connecting nearby qubits, to dynamically manipulate the hopping strengths. 
Together with the dynamical control of the qubits frequencies, we simulate the time-dependent Hamiltonian, enabling the implementation of a cyclic pump. 
During the first half pumping procedure, the system evolves from the topologically trivial regime to the topologically non-trivial regime. At the half pumping period for both diagonal and non-diagonal pumps, the systems evolve to two different types of topological corner states, respectively. 
We experimentally observe the average amount of charge transport 
during the diagonal and non-diagonal pumps, which agree well with the anticipated fractional charge distributions~\cite{wienand2022thouless}. We also experimentally investigate the diagonal HOSPT pump with on-site potential disorder.
Our results open up the avenue for studying HOSPT pumps on various quantum-simulating platforms.

\begin{figure}
    \centering
    \includegraphics[width=0.47\textwidth]{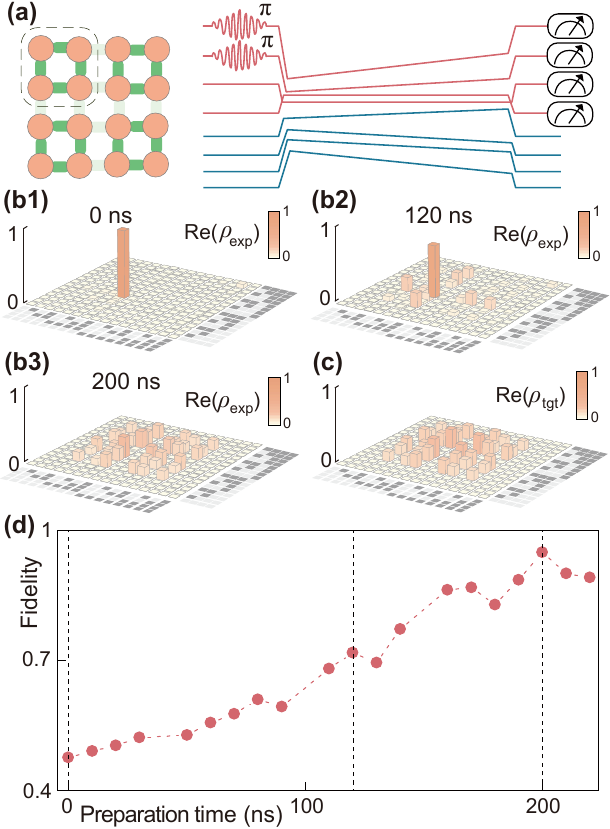}
    \caption{\label{fig:combfig2}
        Initial ground state preparation.
        (a) 16-qubit lattice and pulse sequences on a 4-qubit sub-lattice. Two off-diagonal qubits are excited to $|1\rangle$ with two $\pi $ pulses. Then, the 4-qubit entangled state is prepared by adiabatically controlling the frequencies of both qubits and couplers.
        (b1--b3) 4-qubit density matrices $\rho_{\textrm{exp}}$, measured by quantum state tomography at different evolution times.
        (c) 4-qubit density matrix $\rho_{\textrm{tgt}}$ of the reduced target state in \req{eq:initial_state}.
        (d) Fidelity of the evolving 4-qubit state versus the preparation time, compared with the target state. }
\end{figure}

\emph{Set-up and model.}---On our processor, a flip-chip technique
is employed to integrate 62 superconducting qubits, arranged in a 2D
square lattice, and to incorporate tunable couplers as lattice
bonds~\cite{li2023mapping}, see \rfig{fig:combfig1}(a). In our experiments,
a subset of 16 qubits configured in a 4$\times$4 square lattice is selected,
see \rfig{fig:combfig1}(b). The system can be described as the Bose-Hubbard
model~\cite{you2020higher,bibo2020fractional,grusdt2013topological} with
negative on-site nonlinear interactions with an average value of $-260$~MHz,
which is much larger than the hopping strength.
Thus, the system can be effectively described with hard-core bosons, 
which does not alter the results of our study~\cite{wienand2022thouless,yan2019strongly}. The Hamiltonian under open boundary
conditions (OBC) reads
\begin{equation}
        \hat{H}^{\textrm{OBC}} = -\left[\sum_{x=-D}^{D-1}\sum_{y=-D}^{D}
                (J_{x} \hat{a}_{x,y}^{\dagger}\hat{a}_{x+1,y}+ H.c.)+x \leftrightarrow y \right]
\label{eq:hamiltonianOBC}
\end{equation}
where $\hat{a}^{\dagger}_{x,y}$ ($\hat{a}_{x,y}$) denotes the hard-core bosonic creation (annihilation) operator. 
The subscript $(x,y)$ denotes the coordinate of the qubit, with $x$ and $y$
both varying from $-1.5$ to $1.5$ and $D=1.5$, see \rfig{fig:combfig1}(d).
Here, $J({\xi})$, with $\xi \in \{ x,y\}$, denotes the staggered hopping strength, as
$ J({\xi}) ={J}_0-J$ for $\xi \in \{-1.5, 0.5\}$ and $J({\xi}) =J$ for $\xi=-0.5$,
with $J\in[0,J_0]$.

In our experiments, we demonstrate two types of HOSPT
pumps that transport charge diagonally and non-diagonally, respectively, whose Hamiltonians  read
\begin{align}
\label{eq:Hdiag}
    \hat{H}^{\textrm{diag}} &= \hat{H}^{\textrm{OBC}} + h\sum_{x,y = -D}^{D}
                (-1)^{x+y} \hat{a}_{x,y}^{\dagger}\hat{a}_{x,y}, \\
    \hat{H}^{\textrm{nondiag}} &= \hat{H}^{\textrm{OBC}} - h\sum_{x,y = -D}^{D}
                (-1)^{x+D} \hat{a}_{x,y}^{\dagger}\hat{a}_{x,y}.\label{eq:Hnondiag}
\end{align}
Here, $J$ and $h$  vary periodically with the time $t$ as $J={J}_0\cos \lambda(t)$, $h= {h}_0\sin \lambda(t)$, with $\lambda(t) =\pi -2\pi t/T_0$, with $T_0$ being the pumping period. 
We choose the period as $T_0=500$~ns that is much shorter than the average qubits decoherence times $\overline{T}_1=17.5$~$\mu$s and $\overline{T}_2=2.7$~$\mu$s, see  SM~\cite{suppl}.  
We choose ${J}_0/2\pi=3$~MHz, ${h}_0/2\pi=10$~MHz, $ {J}'_0/2\pi=3$~MHz and ${h}'_0/2\pi=3.5$~MHz for the diagonal and non-diagonal pumps, respectively.

\begin{figure}[t]
    \centering
    \includegraphics[width=0.47\textwidth]{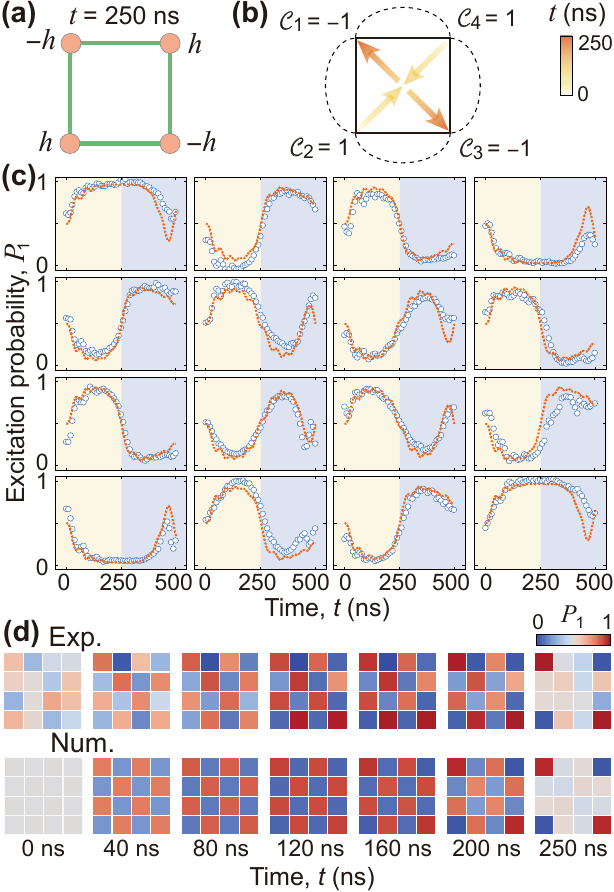}
    \caption{\label{fig:combfig3}
        Diagonal HOSPT pumping.
        (a) Additional on-site potentials for diagonal HOSPT pumping.
        (b) Chern number tuples for diagonal HOSPT pumping, and each Chern number 
        $\mathcal{C}_{1,2,3,4}$  is associated with the charge transport current from the off-diagonal corners to the diagonal corners. 
        (c) Time evolutions of the measured excitation probabilities $P_1$ for all 16 qubits (blue circles) during the 
        full pumping period from $t=0$~ns to $500$~ns, which are compared with the numerical simulations (red dotted curves).
        (d) Measured excitation probabilities $P_1$ for all 16 qubits at different evolution times, which are compared with numerical simulations. At the half pumping period $t=250$~ns, four corner states are observed.
        }
\end{figure}

\begin{figure}[t]
    \centering
    \includegraphics[width=0.47\textwidth]{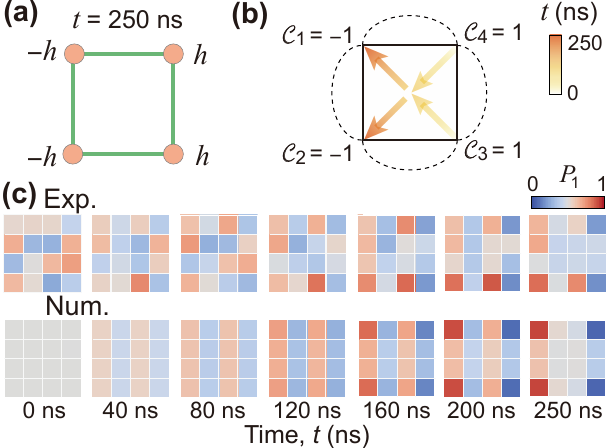}
    \caption{\label{fig:combfig4}
        Non-diagonal HOSPT pumping. 
        (a) Additional on-site potentials for non-diagonal HOSPT pumping.
        (b) Chern number tuples for non-diagonal HOSPT pumping, and each Chern number 
        $\mathcal{C}_{1,2,3,4}$ is associated with the charge transport current from the right-side corners to the left-side corners. 
        (c) Measured excitation probabilities $P_1$ for all 16 qubits at different evolution times,  compared with numerical simulations. At the half pumping period $t=250$~ns, four corner states are observed.
        }
\end{figure}

The system has a $C_4$-symmetry with staggered hopping strengths. 
For the diagonal pump, staggered on-site potentials, $\pm h$, are applied on qubits with the same negative (positive) sign in a diagonal (off-diagonal) direction of the lattice [\rfig{fig:combfig3}(a)]. 
Similarly, for the non-diagonal pump, staggered on-site potentials, $\pm h$, are applied on qubits along the $x$-direction of the lattice, see \rfig{fig:combfig4}(a). Both cases break the $C_4$ symmetry except  $h=0$.
In our experiments, we prepare the initial state as the same ground state
of the Hamiltonians $\hat{H}^{\textrm{diag}}$ and $\hat{H}^{\textrm{nondiag}}$ in Eqs.~(\ref{eq:Hdiag},\ref{eq:Hnondiag}) at $t=0$~ns. As the Hamiltonians evolve, with $J$ and $h$ varying slowly along the closed trajectory as shown in \rfig{fig:combfig1}(c), the system evolves adiabatically to topologically non-trivial phases, manifesting corner localized states at half period $t=T_0/2$, see \rfig{fig:combfig1}(d). 

\emph{Initial ground state preparation.}---The initial state, as the ground state of the superlattice Hamiltonians for half-filling when $t=0$~ns, is topologically trivial. Moreover, 
the 4$\times$4 lattice consists of four independent  2$\times$2 sub-lattices, see 
\rfig{fig:combfig2}(a).
Thus, the initial ground state can be written as a product of four identical 4-qubit entangled states $|\Psi_{\textrm{ini}}\rangle=|\psi_{\textrm{tgt}}\rangle^{\otimes4}$ with
\begin{equation}
    |\psi_{\textrm{tgt}}\rangle =     \frac{1}{\sqrt{8}}
    \left(\left|{}^{0}_{1}{}^0_{1}
    \right\rangle
    +
    \left|{}^{1}_{1}{}^0_{0}
    \right\rangle
    +
    \left|{}^{1}_{0}{}^1_{0}
    \right\rangle
    +
    \left|{}^{0}_{0}{}^1_{1}
    \right\rangle\right)  +\frac{1}{2}\left(
    \left|{}^{0}_{1}{}^1_{0}
    \right\rangle
    +
    \left|{}^{1}_{0}{}^0_{1}
    \right\rangle\right).
    \label{eq:initial_state}
\end{equation}
Notably, the initial state satisfies a particle-hole symmetry for half-filling, resulting in an additional $Z_2$-symmetry in addition to $C_4$-symmetry.

In our experiments, we employ an adiabatic scheme to prepare the required 4-qubit entangled state for each sub-lattice independently. The procedures for all 4 sub-lattices are conducted 
 in parallel, while the couplers connecting different sub-lattices are turned off. 
 For each sub-lattice, we excite two qubits located at the off-diagonal sites, using two $\pi$-pulses transforming $|0\rangle $ to $|1\rangle $,
see \rfig{fig:combfig2}(a). Then, we gradually tune the frequencies of these two qubits from $-21$~MHz to the resonant frequency $0$~MHz, and the hopping strengths between qubits inside each sub-lattice are tuned from $0$~MHz to $6$~MHz adiabatically, see \rfig{fig:combfig2}(a). 
The fidelity of the prepared 4-qubit state $\rho_{\textrm{exp}}$ of each sub-lattice,
compared with the target state $\rho_{\textrm{tgt}}\equiv|\psi_{\textrm{tgt}}\rangle\langle\psi_{\textrm{tgt}}|$, is defined as $F(\rho_{\textrm{tgt}},\rho_{\textrm{exp}}) = {\rm tr}(\rho_{\textrm{tgt}}^{1/2}\rho_{\textrm{exp}}\rho_{\textrm{tgt}}^{1/2})^{1/2}$~\cite{nielsen2010quantum}. The density matrices of $\rho_{\textrm{exp}}$ are obtained by performing quantum state tomography (QST) measurements at different times~\cite{nielsen2010quantum,barends2016digitized}, which are presented in \rfig{fig:combfig2}(b1--b3) and compared with the target state in \rfig{fig:combfig2}(c). The fidelity versus the preparation time is plotted in \rfig{fig:combfig2}(d), and the optimal
 fidelity of the prepared state for each sub-lattice achieves 94.9\%  at 200~ns. More details are presented in SM~\cite{suppl}.

\emph{Diagonal and non-diagonal HOSPT pumping.}---After the initial state preparation,
we simultaneously tune the on-site potentials and the hopping strengths along the cyclic pumping trajectory on the $J$--$h$ plane, through the $Z$-control lines of the qubits and the couplers, respectively. Thus, the systems undergo approximately adiabatic evolutions.
During the first half pumping period, the systems evolve from topologically trivial phases to topologically non-trivial phases, as illustrated in \rfig{fig:combfig1}(c). Furthermore, at the half pumping period $t=250$~ns, the corner localized states appear.

During the full pumping procedure, we measure all 16 qubits in the 
$\{|0\rangle,|1\rangle\}$ basis with 6,000 single-shot readouts every 10~ns, 
and the excitation probabilities of all 16 qubits $P_1$ for the diagonal second-order 
topological pump are plotted in \rfig{fig:combfig3}(c) 
versus the evolution time $t$ and compared with the numerical results. 
At the half pumping period $t=250$~ns, the corner localized states
clearly appear, see Figs.~\ref{fig:combfig3}(d) and \ref{fig:combfig4}(c) for the diagonal and 
non-diagonal pumps, respectively. The experimental results are compared with numerical simulations
using the same experimental parameters.

Due to the bulk-boundary correspondence, a tuple of Chern numbers $\mathcal{C}_{1,2,3,4}$ 
 can be introduced under corner-periodic boundary conditions (CPBCs) \cite{wienand2022thouless}. At the half pumping period, the amount of the transport  charge at each corner $\Delta q_{i} $  is related to the Chern number as
\begin{equation}
\label{eq:qandchern}
    \Delta q_{i} = -{\mathcal{C}_i}/{2},
\end{equation}
which is obtained by measuring the change of the excitation probability $\Delta P^{(i)}_1$ at each corner  $i\in \{1,2,3,4\}$.
Since $\mathcal{C}^{\textrm{diag}}_{1,2,3,4} = -1,1,-1,1$ for the diagonal pump and  $\mathcal{C}^{\textrm{nondiag}}_{1,2,3,4} = -1,-1,1,1$ for the non-diagonal one, the average corner transport charge  ${\Delta q} = \sum_{i=1}^{4}|\Delta q_{i}|/2=1$. 
Here, we obtain that  ${\Delta q} = 0.964$ and $0.555$ for the diagonal and the non-diagonal second-order topological pumps, respectively, which are compared with 0.985
 and 0.836 from the numerical simulation. 
The experimental results for the diagonal pump agree well with the theoretical expectations,
and the imperfections may result from the accuracy of the experimental control on $J$ and $h$,  decoherence, and the inconformity of the qubits on the processor. 
Nevertheless, the non-diagonal HOSPT pump requires a much longer pumping period to fulfill the adiabatic condition, see SM~\cite{suppl} for more details.

\begin{figure}[t]
    \centering
    \includegraphics[width=0.45\textwidth]{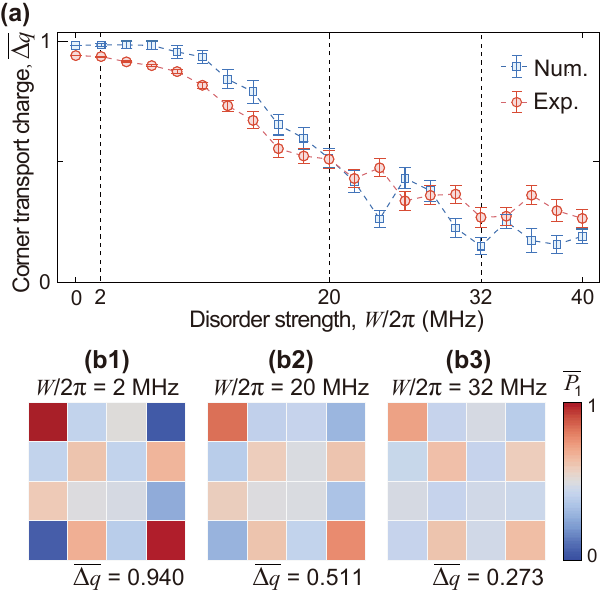}
    \caption{\label{fig:combfig5} (a)
The average amount of transport charge $\overline{\Delta q}$ for diagonal HOSPT pumping versus the disorder strength ranging from $W/2\pi=0$~MHz to $40$~MHz.
(b1--b3) Measured average excitation probabilities $\overline{P_1}$ for all 16 qubits 
at the half 
pumping period with different disorder strengths $W/2\pi=2$~MHz, $20$~MHz, and $32$~MHz with respect to $\overline{\Delta q}=0.940$, 0.511, and 0.273, respectively. 
        }
\end{figure}

\emph{Robustness of HOSPT pumping in the presence of disorder.}---Next, we investigate 
the effects of on-site potential disorder on diagonal HOSPT pumping. 
Here, on-site potential disorder $h=h_0\sin\lambda(t)+\delta h$, we apply on each qubit, following a uniform distribution $\delta h\in [-W,W]$, with $W$ being the disorder strength.
The disorder strength $W/2\pi$ ranges from 0~MHz to 40~MHz, and for each disorder strength, 40 different configurations of the disorder are applied to obtain the average amount of transport charge $\overline{\Delta q}$. Figure~\ref{fig:combfig5}(a) shows
the average amount of transport charge $\overline{\Delta q}$ versus the disorder 
strength $W$, which is compared with numerical simulations using the experimental parameters. The $\overline{\Delta q}$ first decreases
slightly, as $W/2\pi$ increases from 0~MHz, and then decreases
intensely, as $W/2\pi\gtrsim10$~MHz. When $W/2\pi\gtrsim40$~MHz, 
 $\overline{\Delta q}$ tends to be zero, indicating that the quantized 
pump seems to disappear. The effects of on-site disorder on the 
corner-localized states can be observed from the average excitation probabilities $\overline{P_1}$ of all 16 qubits 
at the half pumping period $t=250$~ns, which are shown in \rfig{fig:combfig5}(b1--b3) 
for different disorder strengths.
Our experimental results confirm the robust topological protection of the 
zero-dimension corner state in topological pumping and are in good agreement 
with numerical simulations \cite{lu2023effects}.

\emph{Conclusion and discussion.}---%
We have experimentally demonstrated the diagonal and non-diagonal HOSPT pumps on 
a square-lattice array of 16 superconducting qubits, with accurate dynamical modulations of the hopping strengths and on-site potentials between and on qubits, respectively.  
The HOSPT corner states are observed at the half-pumping period, of which 
the robustness is investigated by introducing on-site disorder.
The ground state of the studied Hamiltonian for half-filling is with a Hilbert space dimension scaling exponentially with the number of qubits, which is entangled and has experimental challenges. The time evolution of the system may involve hard conventional computation, leading to a quantum computational advantage problem~\cite{arute2019quantum}.
Our work would inspire further studies on various HOSPT pumps on different quantum-simulating platforms with a larger system size.


\begin{acknowledgements}
This work was supported by the National Natural Science Foundation of China (Grants Nos.~92265207,
T2121001, 11934018, T2322030, 12122504, 12274142, 92365206, 12104055), the Beijing
Natural Science Foundation (Grant No.~Z200009), the Innovation Program for Quantum Science and Technology
(Grant Nos.~2021ZD0301800, 2021ZD0301802), the Beijing Nova Program
(No.~20220484121), and the Scientific Instrument Developing Project of Chinese Academy of Sciences (Grant
No.~YJKYYQ20200041).

\end{acknowledgements}

\bibliography{main.bib}

\end{document}


\title{Supplementary Materials for ``High-order topological pumping on a superconducting quantum processor''}

\author{Cheng-Lin Deng}
\thanks{These authors contributed equally to this work.}

\affiliation{Institute of Physics, Chinese Academy of Sciences, Beijing 100190, China}
\affiliation{School of Physical Sciences, University of Chinese Academy of Sciences, Beijing 100049, China}

\author{Yu Liu}
\thanks{These authors contributed equally to this work.}

\affiliation{Institute of Physics, Chinese Academy of Sciences, Beijing 100190, China}
\affiliation{School of Physical Sciences, University of Chinese Academy of Sciences, Beijing 100049, China}
\thanks{These authors contributed equally to this work.}
\author{Yu-Ran Zhang}
\thanks{These authors contributed equally to this work.}
\affiliation{School of Physics and Optoelectronics, South China University of Technology, Guangzhou 510640, China}

\author{Xue-Gang Li}
\affiliation{Beijing Academy of Quantum Information Sciences, Beijing 100193, China}

\author{Tao Liu}
\affiliation{School of Physics and Optoelectronics, South China University of Technology, Guangzhou 510640, China}
\author{Chi-Tong Chen}
\affiliation{Institute of Physics, Chinese Academy of Sciences, Beijing 100190, China}
\affiliation{School of Physical Sciences, University of Chinese Academy of Sciences, Beijing 100049, China}

\author{Tong Liu}
\affiliation{Institute of Physics, Chinese Academy of Sciences, Beijing 100190, China}
\affiliation{School of Physical Sciences, University of Chinese Academy of Sciences, Beijing 100049, China}

\author{Cong-Wei Lu}
\affiliation{Department of Physics, Applied Optics Beijing Normal University, Beijing 100875, China}

\author{Yong-Yi Wang}
\affiliation{Institute of Physics, Chinese Academy of Sciences, Beijing 100190, China}
\affiliation{School of Physical Sciences, University of Chinese Academy of Sciences, Beijing 100049, China}

\author{Tian-Ming Li}
\affiliation{Institute of Physics, Chinese Academy of Sciences, Beijing 100190, China}
\affiliation{School of Physical Sciences, University of Chinese Academy of Sciences, Beijing 100049, China}

\author{Cai-Ping Fang}
\affiliation{School of Physical Sciences, University of Chinese Academy of Sciences, Beijing 100049, China}

\author{Si-Yun Zhou}
\affiliation{School of Physical Sciences, University of Chinese Academy of Sciences, Beijing 100049, China}

\author{Jia-Cheng Song}
\affiliation{School of Physical Sciences, University of Chinese Academy of Sciences, Beijing 100049, China}

\author{Yue-Shan Xu}
\affiliation{Institute of Physics, Chinese Academy of Sciences, Beijing 100190, China}
\affiliation{School of Physical Sciences, University of Chinese Academy of Sciences, Beijing 100049, China}

\author{Yang He}
\affiliation{Institute of Physics, Chinese Academy of Sciences, Beijing 100190, China}
\affiliation{School of Physical Sciences, University of Chinese Academy of Sciences, Beijing 100049, China}

\author{Zheng-He Liu}
\affiliation{Institute of Physics, Chinese Academy of Sciences, Beijing 100190, China}
\affiliation{School of Physical Sciences, University of Chinese Academy of Sciences, Beijing 100049, China}

\author{Kai-Xuan Huang}
\affiliation{Beijing Academy of Quantum Information Sciences, Beijing 100193, China}

\author{Zhong-Cheng Xiang}
\affiliation{Institute of Physics, Chinese Academy of Sciences, Beijing 100190, China}
\affiliation{School of Physical Sciences, University of Chinese Academy of Sciences, Beijing 100049, China}
\affiliation{Hefei National Laboratory, Hefei 230088, China}

\author{Jie-Ci Wang}
\affiliation{Department of Physics and Key Laboratory of Low Dimensional\\ Quantum Structures and Quantum Control of Ministry of Education, Hunan Normal University, Changsha, Hunan 410081, China}
 
\author{Dong-Ning Zheng}
\affiliation{Institute of Physics, Chinese Academy of Sciences, Beijing 100190, China}
\affiliation{School of Physical Sciences, University of Chinese Academy of Sciences, Beijing 100049, China}
\affiliation{Songshan Lake Materials Laboratory, Dongguan, Guangdong 523808, China}
\affiliation{CAS Center for Excellence in Topological Quantum Computation, UCAS, Beijing 100190, China, and Mozi Labratory, Zhengzhou 450007, China}
\affiliation{Hefei National Laboratory, Hefei 230088, China}

\author{Guang-Ming Xue}
\affiliation{Beijing Academy of Quantum Information Sciences, Beijing 100193, China}
\affiliation{Hefei National Laboratory, Hefei 230088, China}

\author{Kai Xu}
\email{kaixu@iphy.ac.cn}
\affiliation{Institute of Physics, Chinese Academy of Sciences, Beijing 100190, China}
\affiliation{Beijing Academy of Quantum Information Sciences, Beijing 100193, China}
\affiliation{Songshan Lake Materials Laboratory, Dongguan, Guangdong 523808, China}
\affiliation{CAS Center for Excellence in Topological Quantum Computation, UCAS, Beijing 100190, China, and Mozi Labratory, Zhengzhou 450007, China}
\affiliation{Hefei National Laboratory, Hefei 230088, China}
\author{H. F. Yu}
\email{hfyu@baqis.ac.cn}
\affiliation{Beijing Academy of Quantum Information Sciences, Beijing 100193, China}

\author{Heng Fan}
\email{hfan@iphy.ac.cn}
\affiliation{Institute of Physics, Chinese Academy of Sciences, Beijing 100190, China}
\affiliation{Beijing Academy of Quantum Information Sciences, Beijing 100193, China}
\affiliation{Songshan Lake Materials Laboratory, Dongguan, Guangdong 523808, China}
\affiliation{CAS Center for Excellence in Topological Quantum Computation, UCAS, Beijing 100190, China, and Mozi Labratory, Zhengzhou 450007, China}
\affiliation{Hefei National Laboratory, Hefei 230088, China}

\maketitle

\tableofcontents

\newpage
\clearpage

\section{Device and experimental setup}
\subsection{Device information}
This device contains 62 tunable qubits and 105 tunable couplers integrated with flip-chip technology. Each qubit has an asymmetric superconducting quantum interference device (SQUID) and dispersively couples with a resonator for single-shot readout. The resonator frequency, in the range of $4.1-4.6$~GHz, is far below the qubit idle frequency, around $5.5$~GHz, and every six resonators share a common Purcell filter to suppress the Purcell effect~\cite{purcell1946resonance}. The tunable couplers with higher frequencies ($8$~GHz) are designed to modulate the effective coupling strengths between nearby qubits. More device information can be found in reference~\cite{li2023mapping}, which shared the same device with our experiment.

\subsection{Wiring}
\begin{figure}[h!]
    \centering
    \includegraphics[width=0.71\textwidth]{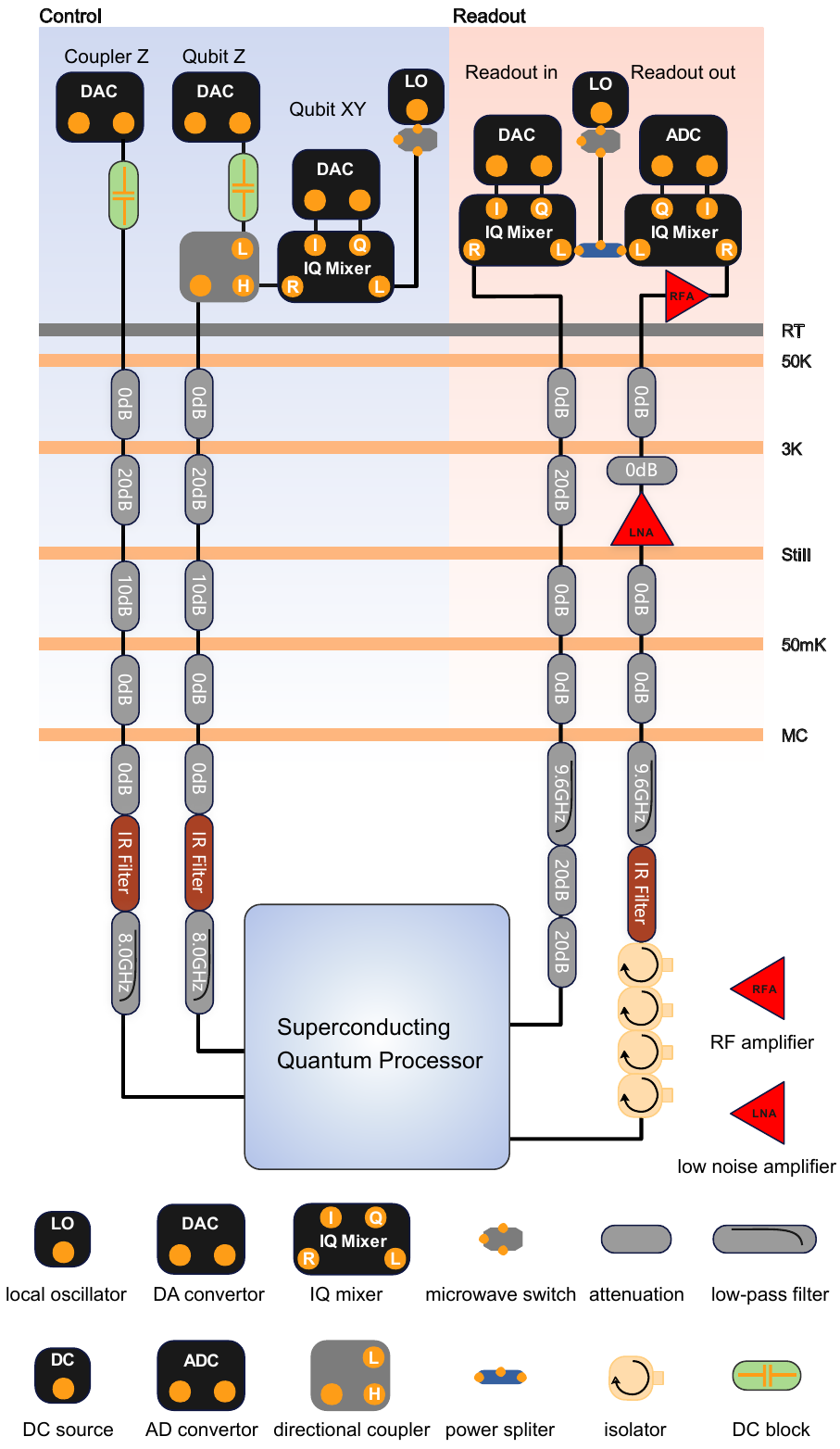}
    \caption{\label{figs:wiringinfo} 
        Schematic diagram of the experimental system and wiring information.
        }
\end{figure}

The typical wiring information is shown in \rfig{figs:wiringinfo}, 
which includes control lines of couplers (for only Z controls), qubits (for XY and Z controls), and readout resonators, from left to right, respectively. 
From up to down, the ambient temperature decreases from room temperature to nearly 12 mK in a BlueFors dilution refrigerator. 
The arrangement of attenuations for all coaxial lines is designed to reduce the thermal noise~\cite{krinner2019engineering}. 
For controlling the qubits and couplers, we combine the high-frequency XY signals and the low-frequency Z biases with directional couplers connected to qubits for state excitation and frequency modulation respectively, while only Z biases are applied to couplers. 
Here, all Z pulses are directly generated by the arbitrary waveform generators (AWGs) and filtered through DC blocks before entering into the dilution refrigerator. XY and readout signals are produced with IQ mixers by mixing the local oscillator (LO) signals generated from microwave source and IQ signals generated from AWGs, known as the upconversion process of IQ mixers, however, the downconversion process, 
which is defined as converting the radio frequency and LO to the intermediate frequency, is used before ADCs collect the experimental data.

\subsection{Effective coupling measurement}
The parameters of qubits are listed in \rtab{tab:chipinfo}. 
\begin{table}[h!]
    \renewcommand\arraystretch{1}
    \tabcolsep=3mm
    \begin{tabular}{lccccccc}
    \hline
    \multicolumn{1}{c}{Qubits} & Idle frequency & Energy relaxation time $T_1$ & Dephasing time $T_2$ & \multicolumn{2}{c}{Readout fidelity} & \multicolumn{2}{c}{Gate fidelity} \\
                               & GHz            & $\mu$s                       & $\mu$s               & $|0\rangle$       & $|1\rangle$      & $X$             & $X/2$           \\ \hline
    $Q_1$                      & 5.284          & 21.988                       & 3.018                & 0.975             & 0.905            & 0.995           & 0.989           \\
    $Q_2$                      & 5.736          & 22.285                       & 2.827                & 0.966             & 0.847            & 0.988           & 0.991           \\
    $Q_3$                      & 5.217          & 22.506                       & 0.956                & 0.991             & 0.896            & 0.967           & 0.995           \\
    $Q_4$                      & 5.761          & 20.516                       & 5.999                & 0.954             & 0.760            & 0.995           & 0.990           \\
    $Q_5$                      & 5.681          & 20.818                       & 2.644                & 0.970             & 0.877            & 0.999           & 0.996           \\
    $Q_6$                      & 5.447          & 25.482                       & 1.435                & 0.953             & 0.900            & 0.989           & 0.986           \\
    $Q_7$                      & 5.313          & 14.196                       & 1.012                & 0.962             & 0.815            & 0.995           & 0.995           \\
    $Q_8$                      & 5.820          & 9.757                        & 2.861                & 0.903             & 0.801            & 0.999           & 0.983           \\
    $Q_9$                      & 5.900          & 11.927                       & 3.067                & 0.986             & 0.869            & 0.999           & 0.991           \\
    $Q_{10}$                   & 5.660          & 15.109                       & 1.841                & 0.943             & 0.883            & 0.990           & 0.999           \\
    $Q_{11}$                   & 5.357          & 18.418                       & 3.653                & 0.948             & 0.871            & 0.998           & 0.986           \\
    $Q_{12}$                   & 5.409          & 4.549                        & 1.627                & 0.952             & 0.698            & 0.998           & 0.999           \\
    $Q_{13}$                   & 5.608          & 16.689                       & 1.621                & 0.785             & 0.783            & 0.999           & 0.999           \\
    $Q_{14}$                   & 5.685          & 17.253                       & 2.109                & 0.952             & 0.896            & 0.999           & 0.999           \\
    $Q_{15}$                   & 5.784          & 20.313                       & 7.268                & 0.925             & 0.842            & 0.988           & 0.995           \\
    $Q_{16}$                   & 5.500          & 17.817                       & 1.034                & 0.952             & 0.880            & 0.989           & 0.988           \\ \hline
    \end{tabular}
    \caption{\label{tab:chipinfo}
        Calibration of qubit parameters. 
        Z bias is applied to shift qubit frequency to the idle point. We optimize the arrangement of all qubits' energy levels to ensure their performance at the idle frequency. Within the standard Bloch-Redfield model of two-level system dynamics, we estimate the energy relaxation time $T_1$ by measuring the exponential decay of the excited qubit and the dephasing time $T_2$ through the Ramsey interferometry~\cite{krantz2019quantum}. We measure the readout fidelities of $\vert 0\rangle$ and $\vert 1\rangle$ state for each qubit, which are used to perform readout corrections. The gate fidelities of $X$ and $X/2$ rotations are characterized by quantum state tomography (QST).
        }
\end{table}

In our experiments, nearby qubits are connected with a coupler (QCQ unit), and we modulate the coupler to vary the hopping strength, $J$, between the nearby qubits. 
Theoretically, $J$ can be quantitatively described as \req{eq:effH}~\cite{yan2018tunable}. 

\begin{equation}
    J = g_{12}+\frac{g_{1c}g_{2c}}{2}\left(\frac{1}{\omega_1-\omega_c}+\frac{1}{\omega_2-\omega_c}\right),
    \label{eq:effH}
\end{equation}
where $g_{12}$ denotes the direct coupling of two qubits, $g_{1c}$ ($g_{2c}$) denotes the coupling strength between the qubit $Q_1$ ($Q_2$) and the coupler. 
Here, $\omega_{1,2}$ and $\omega_c$ denote the frequencies of qubits and the coupler, respectively. 
\begin{figure}[h!]
    \centering
    \includegraphics[width=16 cm]{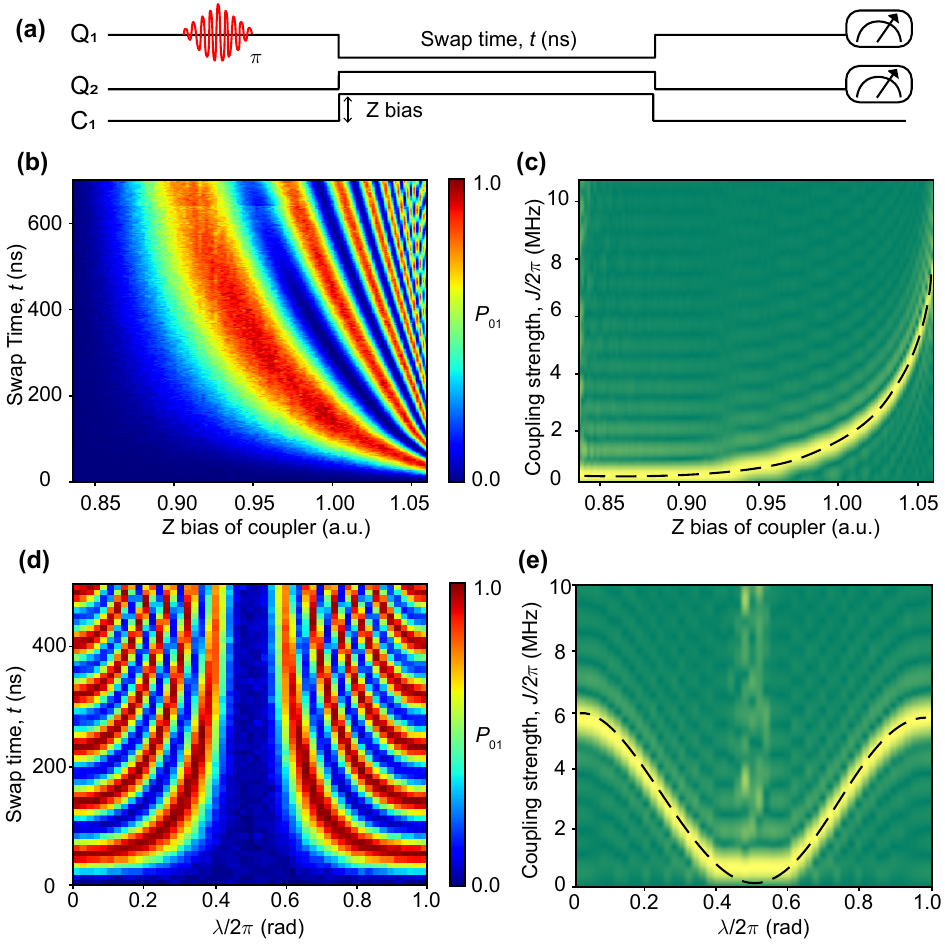}
    \caption{\label{figs:qqswap} 
        Typical measurement of the swap for acquiring the relation between hopping strength $J$ and the Z bias. 
        (a) The pulse sequence is plotted. 
        One of the qubits is excited by a $\pi$ pulse, and then both qubits move to the resonant frequency with the Z biases. The Z bias of the coupler is tuned to modulate the effective coupling strength $J$. 
        (b) With the time evolution, the direct measurement of the probability $P_{01}$ is shown. 
        (c) The swap period, $T_{\textrm{swap}}$, reveals the coupling strength of two qubits. Here, The Fourier transformation is applied to calculate the hopping strength $J$ (dashed curve on the right). We fit this curve using \req{eq:effH} to modulate the control pulse during the pumping procedure.
        (d) Similarly, we measure the swap of nearby qubits to ensure our modulation of hopping strength as $J = J_0 + J_0\cos\lambda(t)$, here, $J_0/2\pi=3$~MHz, $\lambda \in [0,2\pi]$. The direct measurement is plotted in (d) and the hopping strength acquired by the Fourier transformation is shown in (e).
        }
\end{figure} 

Since the hopping strength depends on the Z bias modulation of the coupler, we calibrate every QCQ unit to acquire the relation of Z bias and hopping strength in \rfigs{figs:qqswap}. 
Without losing generality, $J$ can be calculated by measuring the swap period, $T_{\textrm{swap}}$, using $J/2\pi = 1/(2T_{\textrm{swap}})$. Here, we apply the Fourier transformation to the direct measurement plotted in \rfigs{figs:qqswap}~(b) and (d).

\subsection{ac Stark effect compensation}
To demonstrate high-order topological pumping, the accurate modulation of the Z biases of qubits and couplers is a critical factor. In this section, we introduce the experimental method to offset the deviation of the Z bias modulation, which is caused by ac Stark effect. 

\begin{figure}[h!]
    \centering
    \includegraphics[width=16 cm]{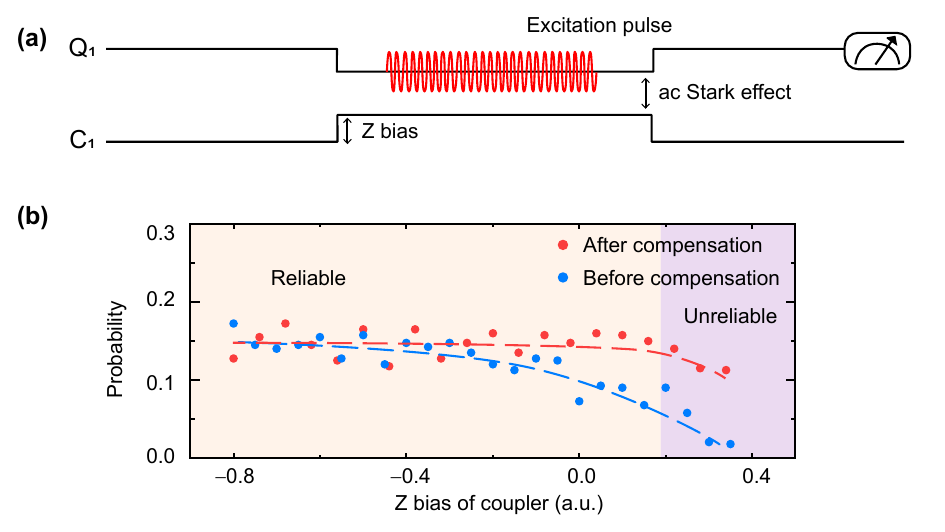}
    \caption{\label{figs:accor} 
        Implementation of the compensation of Z bias, caused by the ac Stark effect. 
        (a) We excite one qubit at a fixed frequency by applying a driving pulse. And then tune the coupler frequency closing to the qubit. We measure the probability of the $\vert 1\rangle$ state to characterize the ac Stark effect.
        (b) The qubit frequency is under repulsion when the coupler approaches, as shown by the blue dashed curve. 
        After compensating the qubit with the extra Z bias, the repulsion is mitigated (red dashed curve). 
        This method does not work when compensating in the unreliable zone, under the detuning condition, where $J \ll \Delta$, is no longer satisfied when $\Delta\to 0$. 
        This limitation requires our experiments to be performed inside the reliable zone. 
        }
\end{figure}

Though the pulse distortions are already corrected when calibrating the single qubit (see the Supplemental Material of Ref.~\cite{xiang2023simulating}), the many-body ac Stark effect still influences our result.
When a coupler is tuned to approach a pair of qubits to acquire enough effective coupling strength, 
the frequency of qubits will shift $J^2/\Delta$, $\Delta = \omega_c-\omega_q$, which deviates the originally scheduled trajectory, see the blue dashed line in \rfigs{figs:accor}.

Thus, we apply the extra compensations of the Z biases on qubits when couplers are tuned. 
This method works well in the reliable zone where our experimental parameters are located (\rfigs{figs:accor}) but shows a slight deviation in the unreliable zone. 
The reason for this deviation is the breakdown of the dispersive condition, $J\ll\Delta$. In addition, the fitting curve of the $J^2/\Delta$ diverges when $\Delta \to 0$, which exacerbates the inaccuracy of this compensation. 

Known from \req{eq:effH}, when the coupler is approaching, the hopping strength turns to be larger, which may fall into the unreliable zone where the compensation of the ac Stark effect does not well. Thus, these limitations indicate that a smaller hopping strength $J$ or a larger on-site potential $h$ could mitigate the inaccuracy caused by the ac Stark effect. 

\section{Model and Hamiltonian}
\subsection{Higher-order diagonal pumping}
The topological properties of higher-order symmetry-protected topological phase (HOSPTs) \cite{benalcazar2017quantized, benalcazar2017electric}, in which the topological boundary states emerge in at least two dimensions lower than the bulk, can be characterized by topological charge pumping \cite{citro2023thouless, wienand2022thouless}. In this work, we focus on higher-order pumping based on the two-dimensional ($2$D) superlattice-Bose-Hubbard model (SL-BHM), which is an interacting $C_4(\times\mathbb{Z}_2)$-symmetric HOSPTs developed in Ref.~\cite{wienand2022thouless}. 
\begin{figure}[h!]
    \centering
    \includegraphics[width=16 cm]{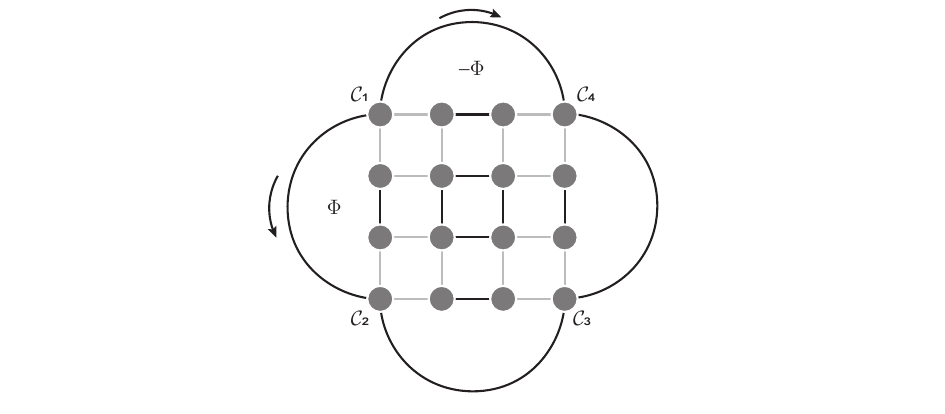}
    \caption{\label{figs:demoCPBC} 
        Superlattice Bose-Hubbard model (SL-BHM) under corner periodic boundary conditions (CPBC). 
        }
\end{figure} 

The SL-BHM contains $N/2\times N/2$ unit cells, as shown in \rfigs{figs:demoCPBC}.  Under open boundary conditions (OBCs), of which the Hamiltonian reads
\begin{equation}\label{eqSLBHM}
    \hat{H}^{\mm{OBC}}=-\left[\sum_{x=-D}^{D-1}\sum_{y=-D}^D(h_x\hat{a}_{x,y}^{\dagger}\hat{a}_{x+1, y}+\mm{H.c.})+x\leftrightarrow y\right]+\frac{U}{2}\!\!\!\sum_{x, y=-D}^D\hat{n}_{x, y}(\hat{n}_{x, y}-1),
\end{equation}
where $D=(N-1)/2$, $\hat{a}_{x, y}^{\dagger}$ ($\hat{a}_{x, y}$) is the bosonic creation (annihilation) operator, $\hat{n}_{x, y} = \hat{a}^\dagger_{x,y}\hat{a}_{x,y}$ is the number operator, $U$ denotes the on-site interaction energy, and $J_x$ $(J_y)$ denotes the staggered hopping strength, with
\begin{equation}
    J_x=
    \begin{cases}
        J_0-J & \text{for}\quad\!\!\! x\in\{-D, -D+2, \cdots, D-1\}\\
        J & \text{for}\quad\!\!\! x\in\{-D+1, -D+3, \cdots, D-2\}.
    \end{cases}
\end{equation}
Diagonal pumping can be realized by driving the Hamiltonian
\begin{equation}
    \hat{H}^{\mm{diag}}=\hat{H}^{\mm{OBC}}+h\!\!\!\sum_{x, y=-D}^D(-1)^{x+y}\hat{n}_{x, y},
\end{equation}
where $h$ denotes the on-site potential, arranged in a cross-diagonal manner. When adiabatically modulating $h$ and $J_x$ ($J_y$) as $h(\lambda)=h_0\sin\lambda$ and $J(\lambda)=J_0(1+\cos\lambda)/2$, with varying $\lambda$ from $0$ to $\pi$, the charges are pumped to corners, yielding four corner-localized fractional charges, as discussed in the main text.

Higher-order pumping is governed by a tuple of higher-order Zak (Berry) phases, which are introduced under the corner periodic boundary conditions (CPBC) that are achieved by adding the corner-connecting links
\begin{equation}
    \hat{H}^C=J(\hat{a}_{c_1}^{\dagger}\hat{a}_{c_2}+\hat{a}_{c_2}^{\dagger}\hat{a}_{c_3}+\hat{a}_{c_3}^{\dagger}\hat{a}_{c_4}+\hat{a}_{c_4}^{\dagger}\hat{a}_{c_1}+\mm{H.c.}),
\end{equation}
to the Hamiltonian Eq.~\eqref{eqSLBHM}. The definition of higher-order Zak phases is based on the idea of adiabatic magnetic flux insertion through two supercells, which can be described by gauge transformations applied only to the corner parts 
\begin{equation}\label{Hc}
    \hat{H}_i^C(\theta)=e^{-i\theta\hat{n}_i}\hat{H}^Ce^{i\theta\hat{n}_i},
\end{equation}
where $\theta$ denotes the flux piercing the supercells outside the bulk, and $\hat{n}_i$ $(i=1, 2, 3, 4)$ is the particle number operator at the $i$-th corner. With the total Hamiltonian $\hat{H}_i(\theta)=\hat{H}^{\mm{OBC}}+\hat{H}_i^C(\theta)$, we can define the higher-order Zak phase as
\begin{equation}
    \gamma_i=i\oint_0^{2\pi}\!\!\mm{d}\theta\bra{\psi_i(\theta)}\partial_{\theta}\ket{\psi_i(\theta)},
\end{equation}
where $\ket{\psi_i(\theta)}$ is the ground state wavefunction of $\hat{H}_i(\theta)$. When evolving along an adiabatic path connecting two HOSPTs, the change of charge at the $i$-th corner, $\Delta q_i$, can be measured by integrating the charge transport current passing through the corner, $\hat{\mathcal{J}}_i=\left.\partial_{\theta}\hat{H}_i(\theta)\right.|_{\theta=0}$. Furthermore, $\Delta q_i$ is proven to be related to the change of the higher-order Zak phase following the expression
\begin{equation}
    \Delta q_i=-\frac{\Delta\gamma_i}{2\pi}.
\end{equation}
During a full pumping cycle, the total amounts of charge are linked to the Chern numbers, $\mathcal{C}_i$, which are obtained as the winding numbers of the Zak phases. Chern numbers are integer, indicating quantized bulk charge transport and read $(-1, 1, -1, 1)$ for higher-order diagonal pumping.

For the first half-Thouless pumping process, as discussed in the main text, with the  $C_4$ symmetry, the change of corner charge, $\Delta q_i$, is given by the Chern number divided by two: 
\begin{equation}
    \Delta q_i =-\frac{\mathcal{C}_i}{2}.
\end{equation}
In practice, we can use twice the average amount of charge transport,
\begin{equation}\label{eq5}
    \Delta q=2\times\frac{1}{4}\sum_{i=1}^4|\Delta q_i|,
\end{equation}
 to characterize the topological properties, and for homogeneously distributed charge density $\Delta q=0$, while $\Delta q=1$ corresponds to the perfect corner-localized density.
 


\subsection{Effects of on-site potential disorder}
\begin{figure}[t]
    \centering
    \includegraphics[width=16 cm]{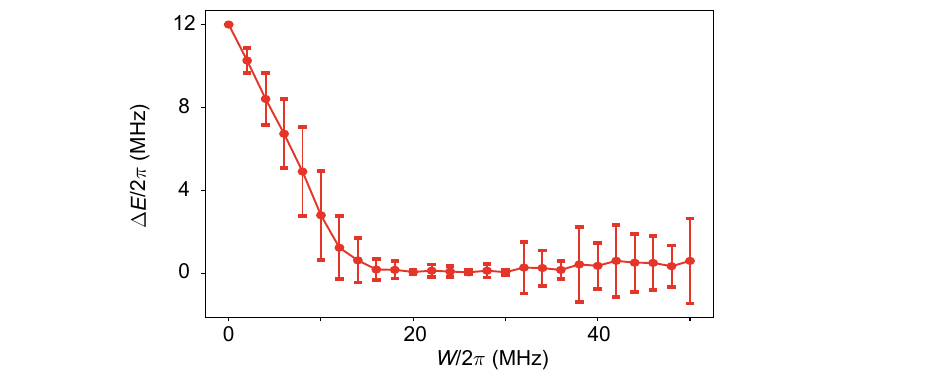}
    \caption{\label{figs:gapwithdisorder} 
        The closing of the energy gap $\Delta E$ of the diagonal pump with the growing disorder strength $W$. The on-site potential disorder $h=h_0\sin\lambda(t)+\delta h$, is applied on each qubit, following the uniform distribution $\delta h \in [-W, W]$. 
        For each disorder strength, 100 repetitions have been simulated randomly based on the BBH model.
        The numerical simulation indicates that the breakdown of $\Delta q$ appears at $W/2\pi \approx 10$~MHz, 
        which agrees well with the experimental validation of robustness in the main text Fig~5. 
        }
\end{figure}

\begin{figure}[p]
    \centering
    \includegraphics[width=14.0 cm]{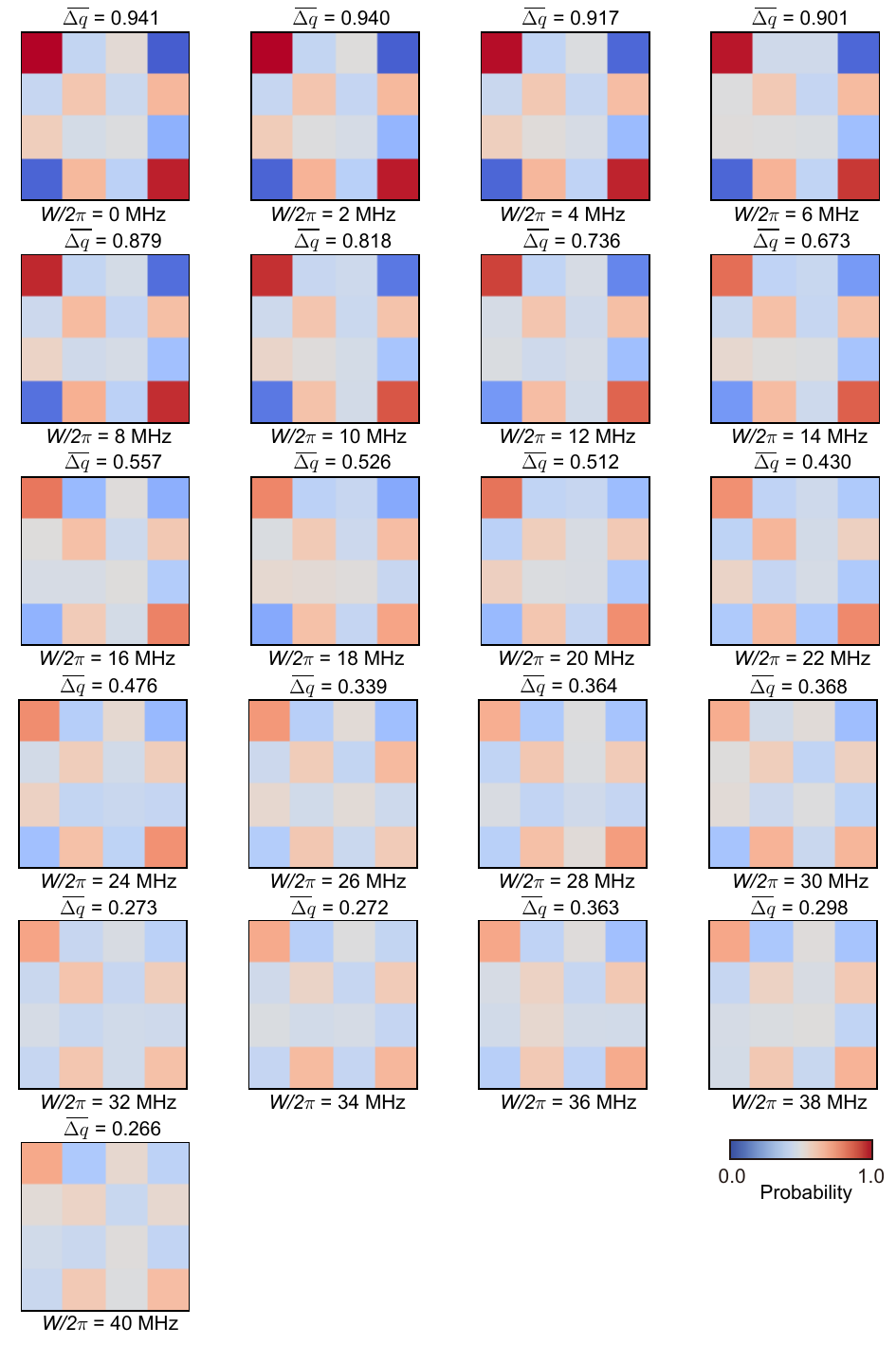}
    \caption{\label{figs:robust_averageface} 
        Zero-dimensional corner states for half-period diagonal pumping in the presence of on-site potential disorder. Experimental results of the diagonal pump against the disorder strength, $W/2\pi =2$~MHz, $20$~MHz, and $32$~MHz, respectively, in the main text Fig~5. More experimental results are shown here. 
        Here, the on-site potential disorder strength $W/2\pi$ is applied from 0~MHz to 40~MHz, stepped by 2~MHz. The results are obtained as the average of the ones from 40 times pumping experiments and each result of the experiment acquires from 6000 single-shot measurements.
        }
\end{figure}

In addition, we investigate the effects of on-site disorder on higher-order diagonal pumping, by introducing on-site potential disorder
\begin{equation}
    \hat{H}^{\mm{disorder}}=W\!\!\sum_{x,y=-D}^D\xi_{x,y}\hat{n}_{x, y},
\end{equation}
where $W$ denotes the disorder strength, and $\xi_{x, y}\in[-1, 1]$ is taken from a uniform distribution for the site $(x, y)$. The total Hamiltonian is written as $\hat{H}=\hat{H}^{\mm{diag}}+\hat{H}^{\mm{dis}}$. Numerically, we calculate $\Delta q$ as a function of disorder strength $W$, based on our experimental parameters, with $h_0/2\pi=10$~MHz, $J_0/2\pi=3$~MHz, $\lambda=\pi-2\pi t/T_0$, and $T_0=0.5$~$\mu s$. The results, as shown in Fig.~5 of the main text, demonstrate that higher-order pumping persists in the regime of sufficiently weak disorder $W/2\pi\lesssim 10$~MHz. For strong disorder $W/2\pi\gtrsim 30$~MHz, $\Delta q\to 0$, the effect of charge transport has disappeared.

Topological effects are stable against perturbations when the energy gap remains intact.
Here, we calculate the minimal many-body instantaneous gap \cite{citro2023thouless, lu2023effects},
\begin{equation}
    \Delta E =\min_{\lambda\in[0, 2\pi]} [E_{N^2/2+1}(\lambda)-E_{N^2/2}(\lambda)],
\end{equation}
with $E_N(\lambda)$ being the $N$-particle ground energy of the $2$D Benalcazar-Bernevig-Hughes (BBH) model \cite{benalcazar2017quantized} under CPBCs. The BBH model is the free fermion analogue of the SL-BHM, and reduces to the BBH model \cite{bibo2020fractional} in the hard-core limit $U\to\infty$ and $J/J_0\in\{0, 1\}$. Figure~\ref{figs:gapwithdisorder} shows that $\Delta E$ closes at $W/2\pi\approx 10$~MHz, in accordance with the breakdown of $\Delta q$. All the experimental results of diagonal pumping with $\Delta q$ under various disorder strengths $W$ are plotted in \rfigs{figs:robust_averageface}.

\subsection{Higher-order non-diagonal pumping} 
In this section, We discuss the experimental challenge of realizing the non-diagonal topological pumping, which can be clarified by the energy gap $\Delta E$.
\begin{figure}[h]
    \centering
    \includegraphics[width=0.87\textwidth]{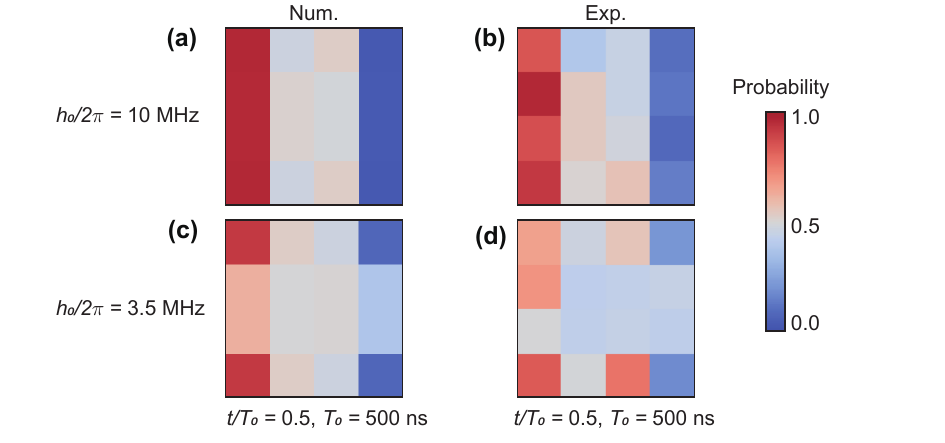}
    \caption{\label{figs:nondiag_expvsnum} 
        Realization of the non-diagonal pump by decreasing the on-site potentials $h_0$.
        (a) The failure of the non-diagonal pump with the on-site amplitude $h_0 = 10$~MHz indicates the inappropriate pumping parameters. Pumping with a half-period, the charge accumulates on both sides of the $4\times 4$ square.
        (b) The experimental result is also consistent with the numerical simulation predictions. Our numerical simulation further shows that the ultra-long pumping period should be exerted when $h_0=10$~MHz, see \rfig{figs:nondiag_evo}. 
        (c) the numerical simulation with $h_0/2\pi = 3.5$~MHz, where the $\Delta q = 0.836$.
        (d) Limited by the decoherence time of this device, we finally perform the non-diagonal pump with $h_0/2\pi=3.5$~MHz and $T_0 = 500$~ns experimentally, with $\Delta q = 0.555$.
        }
\end{figure}

In \rref{wienand2022thouless}, one can rearrange the on-site potentials to change the direction of charge transport, diagonal or non-diagonal. 
Our results realize diagonal transport of the charge within a half cycle of the topological pumping, parameterized by $h_0/2\pi = 10$~MHz, $J_0/2\pi = 3$~MHz, and $T_0 = 500$~ns. 
With these parameters, numerical simulation reveals the results of the non-diagonal pump, which is consistent with our experimental result, see \rfigs{figs:nondiag_expvsnum}. 

\begin{figure}[h!]
    \centering
    \includegraphics[width=16 cm]{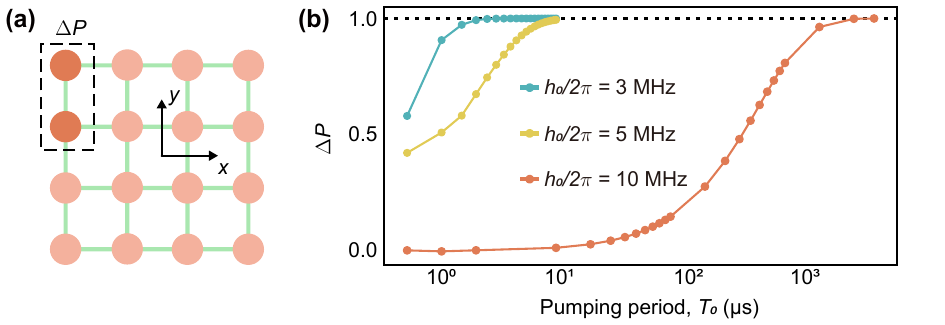}
    \caption{\label{figs:nondiag_evo}
        Numerical simulations of the non-diagonal pump for different on-site potential amplitudes $h_0$. Here, $\Delta P$ denotes the normalized probability difference between one corner and the adjacent along the $y$ axis (a). The pumping period, $T_0$, of achieving the non-diagonal pump is sensitive to the selection of $h_0$. Especially for $h_0/2\pi=10$~MHz (red curve), such an ultra-long $T_0$, here, nearly $5,000$~$\mu$s, is due to the small energy gap, $\Delta E$, as shown in \rfigs{figs:gap_nondiagvsdiag}. The blue curve and the yellow curve demonstrate the feasibility of achieving the non-diagonal pump. Taking into account the ac Stark effect compensation and the finite gap $\Delta E$ of the non-diagonal pump comprehensively, we finally choose the $h_0=3.5$~MHz to demonstrate the non-diagonal pump.
        }
\end{figure}

\begin{figure}[h!]
    \centering
    \includegraphics[width=16 cm]{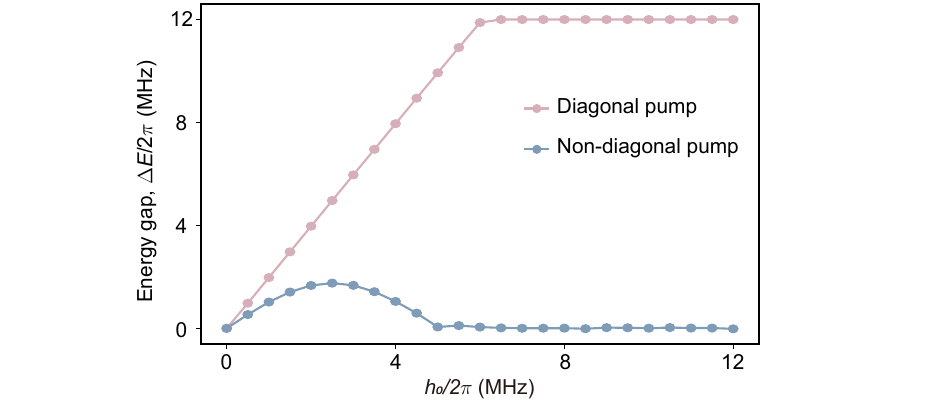}
    \caption{\label{figs:gap_nondiagvsdiag}
        Energy gap under various on-site potentials $h_0$. 
        The finite gap for the diagonal pump remains open but closed as $h_0$ grows for the non-diagonal pump. Thus, we achieve diagonal charge transport with $h_0/2\pi = 10$~MHz, and $T_0=500$~ns, but cannot work on the non-diagonal situation. The energy gap $\Delta E$ achieves the local maximum at around $h_0 = 2.5$~MHz (blue curve). Considering the ac Stark effect compensation, we finally choose the $h_0/2\pi=3.5$~MHz to demonstrate the non-diagonal pump. 
}
\end{figure}

Here, the inappropriate pumping parameters lead to a small energy gap, $\Delta E\to 0 $, shown as \rfigs{figs:gap_nondiagvsdiag}. Our numerical simulation in \rfigs{figs:nondiag_evo} further demonstrates an ultra-long pumping period, $T_0\gtrsim 5,000$~$\mu$s, can achieve the non-diagonal pump, see the red curve in \rfigs{figs:nondiag_evo}. 

Since the ultra-long evolution time of the non-diagonal pump when $h_0/2\pi = 10$~MHz is impractical, we adjust the $h_0/2\pi$ to find a larger energy gap, see the blue curve in \rfigs{figs:gap_nondiagvsdiag}. 
For the numerical evidence, we simulate the non-diagonal pump numerically with $h_0/2\pi = 3.5$~MHz, see \rfigs{figs:nondiag_expvsnum} (c), (d). Limited by the decoherence time, we finally choose the non-diagonal pumping period as $T_0 = 500$~ns, experimentally.

Both experimental analyses and numerical results indicate that the non-diagonal pump requires a more sensitive design of the pumping parameters compared to the diagonal one.

\section{Additional Discussions}
All numerical simulations are performed using QuTip~\cite{johansson2012qutip} (the quantum toolbox in Python). The directly measured qubit probabilities are corrected to eliminate the measurement errors~\cite{xiang2023simulating}.

\clearpage
\bibliography{suppl.bib}